\documentclass[a4paper,11pt,fleqn]{article}
\usepackage{amsfonts,epsfig,latexsym,times}

\newlength{\dinwidth}
\newlength{\dinmargin}
\newcommand\nn{\nonumber}
\newcommand\non{\nonumber\\}
\newcommand{\be}{\begin{equation}}
\newcommand{\ee}{\end{equation}}
\newcommand\ben{\begin{displaymath}}
\newcommand\een{\end{displaymath}}
\newcommand\ba{\begin{eqnarray}}
\newcommand\ea{\end{eqnarray}}

\newcommand{\qed}{\begin{flushright}$\Box$\end{flushright}}

\def\d{\delta}
\def\e{\varepsilon}

\def\g{\gamma}
\def\G{\Gamma}

\def\L{\Lambda}
\def\m{\mu}
\def\n{\nu}
\def\O{\Omega}

\def\r{\rho}
\def\s{\sigma}
\def\S{\Sigma}

\def\th{\theta}

\def\x{\xi}

\def\e{\epsilon}

\def\cI{{\cal I}}
\def\cL{{\cal L}}
\def\cM{{\cal M}}

\def\cP{{\cal P}}
\def\cU{{\cal U}}
\def\cV{{\cal V}}

\def\R{\mathbb{R}}

\def\la{\label}
\def\ci{\cite}

\def\Ref#1{(\ref{#1})}

\def\ft#1#2{{\textstyle {\frac{#1}{#2}} }}
\def\8{\infty}
\def\p{\partial}

\def\tr{{\rm tr}}

\def\gg{\mathfrak{g}}
\def\gk{\mathfrak{k}}
\def\gh{\mathfrak{h}}
\def\GG{\mathbf{G}}
\def\GH{\mathbf{H}}
\def\Coset{\mathbf{G/H}}
\def\i{{\rm i}}
\def\ad{{\rm ad}}

\def\sl2{\mathfrak{sl}(2,\R)}
\def\SL2{SL(2,\R)}
\def\cVh{\hat{\cV}}
\def\rt{\tilde{\rho}}
\def\Dt{\tilde{D}}

\makeatletter
\@addtoreset{equation}{section}
\makeatother

\newtheorem{Remark}{Remark}[section]

\begin{document}
\begin{flushright}
AEI-047\\hep-th/9710210
\end{flushright}

\vspace*{1.3cm}

\noindent
\rule{\linewidth}{0.6mm}
\vspace*{1.2cm}
\renewcommand{\thefootnote}{\fnsymbol{footnote}}
\begin{center}
{\bf\LARGE  Yangian Symmetry in Integrable Quantum Gravity}\bigskip\\
{\bf\large D. Korotkin\footnote{On leave of absence from
    Steklov Mathematical Institute, Fontanka, 27, St.Petersburg 191011,
    Russia.} and H. Samtleben \medskip\\ }
{\large Max-Planck-Institut f\"ur Gravitationsphysik,\\
Albert-Einstein-Institut,\\
Schlaatzweg 1, D-14473 Potsdam, Germany}\smallskip\\ {\small E-mail:
korotkin@aei-potsdam.mpg.de, henning@aei-potsdam.mpg.de\medskip} 
\end{center}
\renewcommand{\thefootnote}{\arabic{footnote}}
\setcounter{footnote}{0}
\vspace*{1.2cm}
\noindent
\rule{\linewidth}{0.6mm}
\vspace*{2cm}
\begin{abstract}
Dimensional reduction of various gravity and supergravity models leads
to effectively two-dimensional field theories described by gravity
coupled $\Coset$ coset space $\s$-models. The transition matrices of
the associated linear system provide a complete set of conserved
charges. Their Poisson algebra is a semi-classical Yangian double
modified by a twist which is a remnant of the underlying coset
structure. The classical Geroch group is generated by the Lie-Poisson
action of these charges. Canonical quantization of the structure
leads to a twisted Yangian double with fixed central extension at a
critical level.

\end{abstract}
\bigskip
\bigskip
\bigskip
\bigskip
\bigskip
\bigskip


\newpage

\section{Introduction}

The last decade has shown the important role of infinite-dimensional
quantum groups in physics, providing a powerful description of the
quantum symmetries of integrable models and field theories. A
prominent example is the Yangian algebra $Y(\mathfrak{\gg})$
associated with a simple finite-dimensional Lie algebra
$\mathfrak{\gg}$. Having turned up already in the early days of the
Quantum Inverse Scattering Method \ci{Skly79,FaSkTa79} this algebra
was rigorously defined within the framework of Hopf algebras by
Drinfeld \ci{Drin85} and later on appeared to underlie many
two-dimensional field theories (see \ci{Bern91,BouSch96} and
references therein). The Yangian algebra $Y(\mathfrak{\gg})$ may be
considered as a deformation of the positive half of a loop algebra
with nontrivial Hopf algebra structure. For
$\mathfrak{g}\!=\!\mathfrak{gl}(N)$ it allows the representation by
matrix entries $T^{ab}(w)$ of $N\!\times\!N$ matrices with exchange
relations
\be\la{YI}
R(v\!-\!w)\stackrel1{T}\!(v)\stackrel2{T}\!(w) ~=~
\stackrel2{T}\!(w)\stackrel1{T}\!(v)\, R(v\!-\!w)\;,
\ee
with a rational $R$-matrix solving the Yang-Baxter equation. A
deformation of the full loop algebra emerges from the Yangian double
construction \ci{Drin86} which has been introduced in quantum field
theory in \ci{LecSmi92,BerLec93}. Like the full loop algebra this
structure admits a central extension \ci{ResSem90,Khor96,IohKoh96}.

In this paper we reveal Yangian symmetries in a class of models of
quantum gravity. Actually, it is the existence of this symmetry in the
corresponding sectors of the classical theory which allows their
complete quantization.  The relevant sectors descend from dimensional
reduction of various gravity and supergravity theories. As effectively
two-dimensional theories they are given by different $\Coset$ coset
space $\s$-models coupled to $2d$ gravity and a dilaton. Among them
(with $\Coset=SL(2,\R)/SO(2))$ is the two Killing vector field
reduction of pure $4d$ Einstein gravity. Its field content is given by
a symmetric $2\!\times\!2$-matrix $g$ with unit determinant, which
parametrizes the $4d$ line element
\be\la{le}
ds^2= e^{2\G(\rho,t)}(-dt^2+ d\rho^2)+ \rho g_{ab}(\rho,t) dx^a
dx^b\;. 
\ee
The accompanying dilaton factor $\r$ is a typical feature of
Kaluza-Klein-like dimensional reductions; here it has already been
identified with a coordinate (the so-called Weyl's canonical
coordinates). Upon further truncation to diagonal $g$ this model
reduces to the Einstein-Rosen waves whose quantization was studied in
\ci{Kuch71,AshPie96a}.  Other examples with higher-dimensional coset
spaces $\Coset$ come from Einstein-Maxwell systems
\ci{KinChi77,Galt95} or maximally extended supergravity
\ci{Juli83}. 

The equations of motion for all these models are given by the Ernst
equation 
\begin{equation}\la{EI}
\p_0(\r\, g^{-1}\p_0 g) 
- \p_1(\r\, g^{-1}\p_1 g) ~=~ 0 \;,
\end{equation}
with a $\GG$-valued matrix $g$ subject to additional coset
restrictions. For $SL(2,\R)/SO(2))$, $g$ alternatively coincides with
the symmetric matrix from \Ref{le} or with a matrix carrying the
dualized potentials of the Ernst picture \ci{Erns68}. Up to the
prefactor of $\rho$ the Ernst equation equals the dynamics of the $2d$
nonlinear $\sigma$-model \ci{LuePoh78,FadRes85,Mack92}. However, it is
exactly this factor which essentially changes many of the properties
of the model and gives rise to several new features.

In particular, it turns out, that in our model the non-ultralocal
contributions of the canonical Poisson brackets do not cause an
obstacle for a well-defined Poisson structure between the generating
functions of the integrals of motion. This allows a reformulation of
the classical model in terms of a complete set of nonlocal conserved
charges $T_\pm(w)$ defined as holomorphic matrix functions in the
upper resp.\ the lower half of the complex plane. 
A consistent and essentially unique canonical quantization of the
complete structure for $\gg\!=\!\mathfrak{sl}(N)$ leads to the
following new algebra of Yangian type:
\ba
R(v\!-\!w)\stackrel1{T_\pm}\!(v)\stackrel2{T_\pm}\!(w) &=&
\stackrel2{T_\pm}\!(w)\stackrel1{T_\pm}\!(v) R(v\!-\!w) \;,\la{YYI}\\
R(v\!-\!w-\i\hbar)\stackrel1{T_-}\!(v)\stackrel2{T_+}\!(w) &=&
\stackrel2{T_+}\!(w)\stackrel1{T_-}\!(v) R^\eta(v\!-\!w+\ft2{N}\i\hbar)
\:\chi(v\!-\!w)\nn\;.
\ea
In contrast to the known centrally extended Yangian double based on
\Ref{YI} the mixed exchange relations here contain an $R$-matrix
$R^\eta$ which is obtained from
$R(v\!-\!w)\in\cU(\mathfrak{g})\!\otimes\cU(\mathfrak{g})$ by
``twisting'' one of the two spaces with the algebra involution $\eta$
which characterizes the underlying coset structure. The values of the
central extension as well as the rescaling factor $\chi(v)$ in
\Ref{YYI} are uniquely fixed. We call this algebra a twisted centrally
extended Yangian double at critical level.

A crucial role in our model is played by the matrix product
\be\la{MI}
\cM(w)=T_+(w)T_-^t(w)\;,
\ee
which in some sense behaves similar to the quantum current introduced
in \ci{ResSem90} for the normal Yangian double. This matrix coincides
with the values of the original field $g$ from \Ref{EI} at
$\r\!=\!0$. This is the symmetry axis if $\r$ is chosen as radial
coordinate; for timelike $\r$ it describes the (cosmological) origin.
Thus, the matrix $\cM(w)$ is localized in the $2d$ spacetime though
at a fixed instant of time it is defined in a highly nonlocal way. It
provides a surprising link between the conserved charges and the
physical fields. In particular, the algebra \Ref{YYI} is compatible
with the symmetry of \Ref{MI}.

The present paper generalizes and gives a detailed account of the
results which were partially announced in \ci{KorSam97a,KorSam97b}. It
is organized as follows.  In Chapter 2 we give the Lagrangian
formulation of the models, derive equations of motion and the
canonical Poisson structure.  Chapter 3 is devoted to the key
calculation of the Poisson algebra between the transition matrices of
the associated linear system. In Chapters 4 and 5 we evaluate this
result to obtain a complete set of conserved charges and their Poisson
algebra after gauge fixing of the dilaton to a spacelike
resp.\ timelike coordinate. Chapter 6 recovers the transitive action of
the Geroch group which in this framework is generated by the
Lie-Poisson action of the classical algebra of charges. The heart of
the paper is Chapter 7 where we present the quantum model for
$\gg\!=\!\mathfrak{sl}(N)$. Quantizing the classical Poisson algebra
we uniquely obtain \Ref{YYI} and show its consistency with all
additional structures. In Chapter 8 we discuss in more detail the
simplest case $\gg\!=\!\mathfrak{sl}(2)$ related to reduced Einstein
gravity \Ref{le}. Chapter 9 briefly summarizes the solved and the
remaining problems.

\section{Canonical Formalism}

Let $\S$ be a two-dimensional Lorentzian world-sheet, parametrized by
coordinates $x^\m$. We introduce light-cone coordinates $x^\pm\equiv
x^0\!\pm\!x^1$. Let $\GG$ be a semisimple Lie group and $\gg$ the
corresponding Lie algebra with basis $t_a$. Denote by $\GH$ the
maximal compact subgroup of $\GG$, characterized as the fixgroup of an
involution $\eta$. Lifting $\eta$ to the algebra gives rise to the
decomposition
\begin{equation}\la{hk}
\gg=\gh\!\oplus\!\gk\qquad\mbox{with}\quad\eta(X)=
\left\{\begin{array}{rl} 
                                X&\mbox{for~} X\in\gh\\
                               -X&\mbox{for~} X\in\gk
                  \end{array}\right. \;,
\end{equation}
which is orthogonal with respect to the Cartan-Killing form.

The physical fields of the model are mappings $\cV(x^\m)$ from $\S$
into the coset space $\Coset$, i.e.\ they are $\GG$-valued and exhibit
the gauge freedom of right $\GH$-multiplication. The currents
$J_\m\equiv J_\m^at_a\equiv\cV^{-1}\p_\m\cV$ allow decomposition
according to \Ref{hk}:  
\begin{equation}\la{PQ}
J_\m = Q_\m+P_\m\;;
\qquad\mbox{with}\quad Q_\m\in\gh\,,\enspace P_\m\in\gk\;,
\end{equation}
such that the gauge transformations take the form
\begin{equation}\la{gauge}
Q_\m~\mapsto~ h^{-1}Q_\m h +  h^{-1}\p_\m h\;,\quad 
P_\m~\mapsto~ h^{-1}P_\m h\;,
\end{equation}
with $h\!=\!h(x^\m)\in \GH$.

Let us establish the canonical setting. The dimensional reduction of
gravity and supergravity theories has already and often been described
in the literature (see e.g.\ \ci{Juli83,BrMaGi88,Nico91}). Hence, we
here restrict to starting from the effectively two-dimensional reduced
theory given by $2d$ dilaton gravity coupled to different nonlinear
$\Coset$ coset space $\s$-models corresponding to different original
theories.  The remaining (reduced and gauge-fixed) Lagrangian is of
the form
\begin{equation}\la{L}
\cL = {\textstyle \frac12}\,\r\;\tr \left(P_\m P^\m\,\right) =
{\textstyle \frac12}\,\r\;\tr \left(P_0^2-P_1^2\right)\;,
\end{equation}
In addition to the coset currents $P_\m$ from \Ref{PQ} there appears a
dilaton field $\r$ related to the compactified part of the former
higher-dimensional metric, which obeys the $2d$ Laplace equation
\begin{equation}\la{rho}
\Box\,\rho=0\;.
\end{equation}
At this stage of reduction $\r$ is no longer a canonical field, but
has already been gauge-fixed to a particular solution of \Ref{rho} on
the world-sheet $\S$. Define its dual $\rt$ by $\p_\m\rt =
-\e_{\m\n}\p^\n\r$.\footnote{We use the convention
$\e_{01}=\e^{10}=1$ for the antisymmetric $\e$-symbol and
$\eta_{\m\nu}=\mbox{diag}(1,-1)$ for the $2d$ Minkowski metric.} In
the sequel, we will often make use of the Weyl gauge choice,
i.e.\ identify $\rho, \rt$ with the world-sheet coordinates $x^\m$.

The Lagrangian \Ref{L} resembles the one of the principal chiral field
model (PCM) \ci{LuePoh78,FadRes85} with the group $\GG$ of the PCM
replaced by the coset manifold $\Coset$ and arising of the additional
dilaton field $\r$ which is in fact responsible for all the
differences that will appear in contrast to the PCM already on the
classical level. Nevertheless, we may introduce the canonical
framework in an analogous way. Treat the current $J_1(x)$ as canonical
coordinates; its time derivative is expressed via the condition of
vanishing curvature:
\ben
\p_0 J_1 = \p_1 J_0 + [J_1,J_0] \equiv \nabla_1 J_0\;.
\een
Note, that the operator $\nabla_1$ is antisymmetric with respect to the
scalar product $\left(\tr\int\!dx^1\right)$. The action thus reads
\ba
\ft12 \int  \rho\: \tr \left(P_\m P^\m\,\right) \,dx^0 dx^1  &=& 
 \ft12 \int  \rho\: \tr \left(P_0 \nabla_1^{-1}(\p_0 J_1) -
P_1^2\right) \,dx^0 dx^1 \non 
&=& -\ft12\int \tr \left((\p_0 J_1) \nabla_1^{-1}(\rho\, P_0) -
\rho P_1^2\right) \,dx^0 dx^1 \;.\nn 
\ea
Introduce the corresponding momenta $\pi_J \equiv \pi_Q\!+\!\pi_P$
with canonical Poisson structure 
\begin{equation}\la{PB}
\left\{J_1^a(x),\pi_J^b(y)\right\} 
= \d^{ab}\,\d(x\!-y)\;,\footnotemark
\end{equation}\footnotetext{Here and in the following we denote for
simplicity the spatial coordinate $x^1$ by $x$ only and the timelike
coordinate $x^0$ by $t$.}  at equal times. From the action it follows
that
\ben
\rho\, P_0 = -\nabla_1 \pi_J = -\p_1\pi_J - [J_1,\pi_J]\;,
\een
which according to \Ref{hk} gives rise to
\ba 
\rho\, P_0 &=& -\p_1\pi_P - [Q_1,\pi_P] - [P_1,\pi_Q]\;,\la{c1} \\
0 &=& -\p_1\pi_Q - [Q_1,\pi_Q] - [P_1,\pi_P] ~\equiv~ \phi \;.\la{c2}
\ea
The first equation expresses a part of the coordinates' time
derivatives in terms of the canonical momenta; the second equation
forms a set of (first-class) constraints, that exactly generate the
gauge transformations \Ref{gauge}.

For further calculations, we switch to the index-free tensor
notation. Denote for some matrix $X$:
\ben
\stackrel{1}{X}\;\equiv X\otimes I\qquad\mbox{and}\qquad
\stackrel{2}{X}\;\equiv I\otimes X\;.\,\footnotemark
\een
\footnotetext{In components this takes the form
  $(X\otimes I)^{ab,cd} \equiv X^{ab}\d^{cd}$
  and $(I\otimes X)^{ab,cd} \equiv
  X^{cd}\d^{ab}\;$.} 
Define accordingly the following matrix notation of Poisson brackets
\ci{FadTak87}:  
\begin{equation}
\Big\{\stackrel1{A}\;,\,\stackrel2{B}\Big\}^{ab,cd} \equiv
\{A^{ab},B^{cd}\}\;,
\end{equation}
for matrices $A^{ab}$, $B^{cd}$. Let $\O_\gg\!\equiv\!  t_a\!\otimes
t^a$ be the Casimir element of $\gg$, which due to orthogonality of
\Ref{hk} allows the decomposition $\O_\gg =\O_\gh+\O_\gk$.  The
canonical brackets \Ref{PB} in this notation become
\ben
\left\{\stackrel1{Q}_1\!(x)\;,\,\stackrel2{\pi}_Q\!(y)\right\} 
= \O_\gh\, \d(x\!-y)\;,\qquad
\left\{\stackrel1{P}_1\!(x)\;,\,\stackrel2{\pi}_P\!(y)\right\} 
=\O_\gk\, \d(x\!-y) \;.
\een
Equ.\ \Ref{c1} now yields the Poisson brackets for the
physical fields:
\ba
\left\{\stackrel1{P}_0\!(x)\;,\,\stackrel2\cV(y)\right\} &=&
-\frac1{\r(x)}\,\stackrel2\cV\!(x)\; \O_\gk\; \d(x\!-y) \;,\la{PBf}\\
\left\{\stackrel1{P}_0\!(x)\;,\,\stackrel2{Q}_1\!(y)\right\} &=&
 \frac1{\r(x)}\;\left[\,\O_\gk\;,\,
\stackrel2{P}_1\!(x)\,\right]\d(x\!-y) 
\;,\non
\left\{\stackrel1{P}_0\!(x)\;,\,\stackrel2{P}_1\!(y)\right\} &=&
 \frac1{\r(x)}\;\left[\,\O_\gk\;,\,
\stackrel2{Q}_1\!(x)\,\right]\d(x\!-y) 
+ \frac1{\r(x)}\;\O_\gk\;\p_x\d(x\!-y)\;, \non
\left\{\stackrel1{P}_0\!(x)\;,\,\stackrel2{P}_0\!(y)\right\} &=&  
 \frac1{\r(x)}\;\left[\,\O_\gh\;,\,
\stackrel2{\phi}\!(x)\,\right]\d(x\!-y) ~\approx~ 0\;.\nn
\ea
Moreover, it follows from \Ref{c2} that the constraints $\phi(x)$
indeed build a closed algebra
\begin{equation}\la{con}
\left\{\stackrel1{\phi}\!(x)\;,\,\stackrel2{\phi}\!(y)\right\} ~=~
 \frac1{\r(x)}\;\left[\,\O_\gh\;,\,
\stackrel2{\phi}\!(x)\,\right]\d(x\!-y) \;,
\end{equation}
and infinitesimally generate the gauge transformations
\Ref{gauge}. They obviously leave the Lagrangian \Ref{L} invariant.
\begin{Remark}\rm
The Poisson brackets \Ref{PBf} are the canonical Poisson brackets
derived from the Lagrangian \Ref{L} which in turn descends from a
consistent dimensional reduction of the higher-dimensio\-nal Lagrangian
of the original theory. Thus, this Poisson structure comes from proper
reduction of the symplectic structure of the original theory.
\end{Remark}

Another important feature to note about the Poisson brackets \Ref{PBf}
is the appearance of a non-ultralocal term in the third equation. In
the known flat space integrable models the presence of such a term is
a good indicator for some breakdown of the conventional techniques at
later stage (see e.g.\ \ci{VeEiMa84} for exploring the fatal
consequences of the non-ultralocal term in the PCM). However, for our
model we will see that this term shows a surprisingly good behavior
and in fact supports the entire further treatment.

Finally, we can discuss the dynamics of the model. The equations of
motion derived from \Ref{L} are 
\begin{equation}\la{eqm}
D^\m(\r\, P_\m) = D_0(\r\, P_0) - D_1(\r\, P_1) = 0\;,
\end{equation} 
with the covariant derivative $D_\m P_\nu\!\equiv\!\p_\m P_\nu
+[Q_\m\,,P_\nu]$. In terms of the field $g\equiv\cV\eta(\cV)^{-1}
\in \GG$ this becomes the more familiar form of the Ernst equation
\ci{Erns68}
\begin{equation}\la{Ernst}
\p^\m(\r\, g^{-1}\p_\m g) ~=~ \p_0(\r\, g^{-1}\p_0 g) 
- \p_1(\r\, g^{-1}\p_1 g) ~=~ 0 \;.
\end{equation}

The Hamiltonian of the model comes out to be
\begin{equation}\la{H}
H = \ft12 \int \rho\: \tr \left(P_0^2 + P_1^2 \right) \, dx \;.
\end{equation}
There are two points to mention about the Hamiltonian dynamics
here. Note first that $H$ does not govern the possible explicit
time-dependence of $\r$. This is due to the fact that in course of the
reduction leading to \Ref{L} the system has become nonautonomous due
to the particular gauge-fixing of the dilaton $\r$ to a function with
explicit time-dependence. It implies in particular that $H$ generates
the equations of motion only for the fields $(\r P_0)$ and
$P_1$. Secondly, $H$ generates the dynamics of these fields only up to
gauge transformations \Ref{gauge} as can be explicitly checked. This
corresponds to the fact that as a Hamiltonian of a constrained system
$H$ is determined only up to a linear combination of the arising
first-class constraints \Ref{con}. 

\section{Poisson Algebra of Transition Matrices}\la{transition}

In this and the following chapters we will exploit the integrability
of the model to define and explore the transition matrices to be used
as the fundamental objects in the sequel.  The model \Ref{L} is
integrable in the sense that it possesses a linear system
\ci{BelZak78,Mais78}; i.e.\ the equations of motion \Ref{eqm} appear as
integrability conditions of the following linear system of
differential equations:
\begin{equation}\la{ls}
\p_\m \cVh(x,t,\g) = \cVh(x,t,\g) L_\m(x,t,\g)\;,
\end{equation}
with
\ben
L_\m(x,t,\g) = Q_\m + \frac{1+\g^2}{1-\g^2}\,P_\m +
\frac{2\g}{1-\g^2}\,\e_{\m\n}P^{\n} \;,
\een
and
\ben
\g(x,t,w) = \frac1{\r}\left(w+\rt-\sqrt{(w+\rt)^2-\r^2}\;\right)\;.
\een

The PCM admits a similar linear system with constant spectral
parameter $\g$ \ci{Pohl76,ZakMik78}. The difference here which is
essential for the entire following treatment is the coordinate
dependence of $\g$ and its interplay with the underlying constant
parameter $w$. Some useful and illustrative formulas are collected in
Appendix \ref{spectral}.

The existence of the linear system allows the construction of the
transition matrices
\ben
T(x,y,t,w)~\equiv~\cVh^{-1}(x,t,\g(x,t,w))\:\cVh(y,t,\g(y,t,w))\;.
\een 
They satisfy: 
\ba
T(x,x,t,w)&=& I \;,\la{TT}\\
\p_x T(x,y,t,w) &=& -L_1(x,t,\g(x,t)) T(x,y,t,w) \;,\non
\p_y T(x,y,t,w) &=& T(x,y,t,w) L_1(y,t,\g(y,t)) \;,\non
\p_t T(x,y,t,w) &=& -L_0(x,t,\g(x,t)) T(x,y,t,w)
+ T(x,y,t,w) L_0(y,t,\g(y,t)) \;.\nn
\ea
Like the spectral parameter $\g$, the transition matrices also live on
a twofold covering of the complex $w$-plane. Transition between the
two sheets is performed by the involution $\eta$. Until explicitly
stated, we shall in the following always consider the sheet with $\g\in
D_+\cup D_-$ inside of the unit circle (cf. Appendix \ref{spectral}).

Note, that $T(x,y,t,w)$ is uniquely determined by the first three
equations of \Ref{TT}; its time-dependence is a consequence of the
equations of motion \Ref{eqm}. If the physical currents and thus $L_0$
vanish sufficiently fast at spatial infinity, \Ref{TT} already shows,
that the transition matrices connecting the spatial boundaries become
integrals of motion. They shall in fact play the main role in the
sequel. The rest of this chapter will be spent to calculate their
Poisson algebra.

Let $T(x,y,v)$ and $T(x',y',w)$ be the transition matrices with
spectral parameters $v$ and $w$ respectively and pairwise distinct
endpoints $x, y$ and $x', y'$.\footnote{For clearness we drop the
(coinciding) argument $t$ for the rest of this chapter.} Equ.\
\Ref{TT} implies the well-known formula \ci{FadTak87}
\ba
\left\{\stackrel1{T}\!(x,y,v)\;,\,\stackrel2{T}\!(x',y',w)\right\}
&=&
\int_x^y\!dz\int_{x'}^{y'}\!dz'\enspace
\Big(\!\stackrel1{T}\!(x,z,v)\stackrel2{T}\!(x',z',w)\Big) \la{n}\\
&& \hspace{-1.9em}
\left\{\stackrel1{L_1}\!(z,\g(z,v))\;,
\stackrel2{L_1}\!(z',\g(z',w))\right\}
\Big(\!\stackrel1{T}\!(z,y,v)\stackrel2{T}\!(z',y',w)\Big)\;.\nn
\ea

Due to the underlying coset structure of the model, it is already far
from obvious, that the Poisson algebra of the connection $L_1$ of the
linear system \Ref{ls} is of a closed form. However, this comes out to
be true on the constraint surface \Ref{c2}: 
\ba
\left\{\stackrel1{L}_1\!(x,\g_1)\;\;,\;
\stackrel2{L}_1\!(y,\g_2)\right\} &=& 
-\frac{2\g_1\g_2}{\r(\g_1-\g_2)(1-\g_1\g_2)}
\left[\O_\gh\;,\; \stackrel1{L}_1\!(\g_1)
+\!\stackrel2{L}_1\!(\g_2)\right]\d(x-y)\non
&& {}-\frac{2\g_2^2(1-\g_1^2)}{\r(1-\g_2^2)(\g_1-\g_2)(1-\g_1\g_2)}
\left[\O_\gk\;,\;\stackrel1{L}_1\!(\g_1)\right]\d(x-y) \non
&&{}- \frac{2\g_1^2(1-\g_2^2)}{\r(1-\g_1^2)(\g_1-\g_2)(1-\g_1\g_2)}
\left[\O_\gk\;,\;\stackrel2{L}_1\!(\g_2) \right]\d(x-y)\non
&&{}-\frac{2\,\O_\gk }{(1\!-\!\g_1^2)(1\!-\!\g_2^2)}
\left(\frac{\g_1(1\!+\!\g_2^2)}{\r(x)}+
\frac{\g_2(1\!+\!\g_1^2)}{\r(y)}\right)
\p_x\d(x-y)\;,\nn
\ea
with $\g_1\!\equiv\!\g(x,v)$, $\g_2\!\equiv\!\g(y,w)\;.$
Inserting this into \Ref{n} finally leads to the following result:
\begin{equation}\la{PBT}
\left\{\stackrel1{T}\!(x,y,v)\;,\,\stackrel2{T}\!(x',y',w)\right\} ~=
\end{equation}
\vspace*{-1.3em}
\ba
&&\hspace{1em}{}\frac1{v-w}\;\times\;\;\left\{\;\;
\th(x,x',y)\;\Big(\!\stackrel1{T}\!(x,x',v)\;\;\O_\gh\;
\stackrel1{T}\!(x',y,v)\!\stackrel2{T}\!(x',y',w)\Big)\right.\non
&&\hspace{6em}
{}+\th(x',x,y')\;
\Big(\!\stackrel2{T}\!(x',x,w)\;\;\O_\gh\;
\stackrel1{T}\!(x,y,v)\!\stackrel2{T}\!(x,y',w)\Big)\non
&&\hspace{6em}{}-\th(x,y',y)\;
\Big(\!\stackrel1{T}\!(x,y',v)\!\stackrel2{T}\!(x',y',w)\;\;\O_\gh\;
\stackrel1{T}\!(x,y,v)\Big)\non
&&\hspace{6em}{}\left.-\th(x',y,y')\;
\Big(\!\stackrel1{T}\!(x,y,v)\!\stackrel2{T}\!(x',y,w)\;\;\O_\gh\;
\stackrel2{T}\!(y,y',w)\Big)\;\right\}\non
&&\non
&&{}+\frac{\th(x,x',y)}{v-w}\;
\Big(\!\stackrel1{T}\!(x,x',v)\;\;\O_\gk\;
\stackrel1{T}\!(x',y,v)\!\stackrel2{T}\!(x',y',w)\Big)\;
\frac{\g(x',v)(1\!-\!\g^2(x',w))}{\g(x',w)(1\!-\!\g^2(x',v))}\non
&&{}+\frac{\th(x',x,y')}{v-w}\;
\Big(\!\stackrel2{T}\!(x',x,w)\;\;\O_\gk\;
\stackrel1{T}\!(x,y,v)\!\stackrel2{T}\!(x,y',w)\Big)
\;\frac{\g(x,w)(1\!-\!\g^2(x,v))}{\g(x,v)(1\!-\!\g^2(x,w))}
\non
&&{}-\frac{\th(x,y',y)}{v-w}\;
\Big(\!\stackrel1{T}\!(x,y',v)\!\stackrel2{T}\!(x',y',w)\;\;\O_\gk\;
\stackrel1{T}\!(x,y,v)\Big)
\;\frac{\g(y',v)(1\!-\!\g^2(y',w))}{\g(y',w)(1\!-\!\g^2(y',v))}\non
&&{}-\frac{\th(x',y,y')}{v-w}\;
\Big(\!\stackrel1{T}\!(x,y,v)\!\stackrel2{T}\!(x',y,w)\;\;\O_\gk\;
\stackrel2{T}\!(y,y',w)\Big)
\;\frac{\g(y,w)(1\!-\!\g^2(y,v))}{\g(y,v)(1\!-\!\g^2(y,w))}\;\;,\nn
\ea
where we have made use of the abbreviation: 
\ben
\th(x,y,z)=\left\{\begin{array}{ll} 
                                1&\mbox{for~} x<y<z\\
                                0&\mbox{else}
                  \end{array}\right.\;.
\een 

Although during the calculation several additional terms arise due to
the explicit coordinate dependence of the spectral parameter $\g$, the
result takes a similar form similar as in the PCM.  At first sight,
we thus face the same fatal problem: With distinct endpoints $x, x', y,
y'$ the algebra is uniquely and well defined, satisfying in particular
antisymmetry and Jacobi identities. For coinciding endpoints on the
other hand equ.\ \Ref{PBT} obviously gives an ambiguous answer,
e.g. for $x\!=\!x'$ due to the different coefficients of
$\th(x',x,y')$ and $\th(x,x',y)$). In the PCM this further leads to
ambiguities in the limit $x,x'\rightarrow-\8,\,y,y'\rightarrow\8$ with
no possibility to restore it in accordance with antisymmetry and
validity of Jacobi identities \ci{LuePoh78,VeEiMa84}.\footnote{Several
  procedures have been suggested to nevertheless give sense to the
  classical Poisson algebra of the PCM \ci{FadRes85,DuNiNi90,Mack92}.}

In our model however, the coordinate dependence of the spectral
parameter changes the situation drastically. Note first, that the part
of \Ref{PBT} related to the subalgebra $\gh$ and thus revealing the
underlying coset structure (the first four lines) allows a unique
limit to coinciding endpoints. The remaining terms invoke the spectral
parameters $v$ and $w$ in combinations of the form

\begin{equation}\la{amb}
\frac{\g(x,w)(1-\g^2(x,v))}{\g(x,v)(1-\g^2(x,w))}
~=~ \sqrt{\frac{(v+\rt)^2-\r^2}{(w+\rt)^2-\r^2}}\;.
\end{equation}
This shows the interesting consequence of the explicit coordinate
dependence of $\g$: if at spatial infinity the fields $\r$ or $\rt$
diverge --- like they do for the choice of Weyl's canonical
coordinates --- all these terms tend to $\pm1$. In particular, they
become independent of $v$ and $w$, thus \Ref{PBT} possesses a well
defined and unique limit for coinciding endpoints at spatial
infinity. As a result, the Poisson algebra \Ref{PBT} for these
transition matrices takes a treatable form that is related to the
well-known Yangian algebra \ci{Drin85}. We shall study this in detail
for fixed choices of $\r$ and $\rt$ in the subsequent chapters.

\section{Spacelike Dilaton}

Assuming the vector field $\p_\m\r$ to be globally spacelike, we now
identify $\r$ with the radial coordinate $x\!=\!r\in[0,\8[$. This is a
usual choice of coordinates for describing cylindrically symmetric
gravitational waves \ci{Kuch71,AshPie96a}. The dual field $\rt$ then
is identified with the time $t$.

Consider the object
\begin{equation}\la{V1}
\cVh_{1}(r,t,\g(w))\equiv \cV(r\!=\!0,t)T(0,r,t,w)\;,
\end{equation}
with $\g\in D_+\cup D_-$ inside of the unit circle. 

According to \Ref{TT} and \Ref{limits}, this is a solution of the
linear system \Ref{ls}. A closer look at the properties of the
spectral parameter (cf. Appendix \ref{spectral}) shows, that it is the
unique solution which is holomorphic inside of the unit circle in the
$\g$-plane.\footnote{This normalization for the solution of the linear
  system has e.g.\ been chosen in \ci{BreMai87}.} 

Let us assume, that the physical currents $J_\m$ fall off sufficiently
fast at spatial infinity $r\!\rightarrow\!\8$. In this limit,
$\cVh_{1}$ then becomes $t$-independent. As a function of $w$ it
becomes discontinuous along the real $w$-axis (since the branch cut
blows up and cuts the plane into two halves). We denote them by
\begin{equation}\la{Tpm}
T_\pm(w)~\equiv~\overline{T_\mp(\bar{w})}~\equiv~
\cV(r\!=\!0,t)T(0,\8,t,w)\qquad\mbox{for $\g(w)\in D_\pm$}\;.  
\end{equation}

These constants of motion $T_\pm(w)$ will provide the new variables of
the model. 
As functions of the constant spectral parameter $w$ they are
holomorphic in the upper resp. lower half of the complex plane.
We can still bring them into a more illustrative form.
Starting from 
\ben
T_\pm(w) = \cV(r\!=\!0,t)\;\,\cP\exp\int_0^\8 \left(Q_1 + 
\frac{1+\g^2}{1-\g^2}\,P_1 -
\frac{2\g}{1-\g^2}\,P_0\right) dr \;,
\een
the $t$-independence may be exploited to calculate this expression for
real $w$ at the specific value $t=-w$ (assuming regularity of the
currents): 
\begin{equation}\la{rc}
T_\pm(w) ~=~ \cV(r\!=\!0,t\!=\!-w)\;\,\cP\exp\int_0^\8 
\Big(Q_1(r,-w)\pm\i P_0(r,-w)\Big) dr \;.
\end{equation} 
Thus, on the real $w$-axis $T_\pm(w)$ naturally factorizes into the
product of a real and a compact part. It follows, that the matrix
product 
\begin{equation}\la{MTT}
\cM(w) ~\equiv~ \lim_{\e\rightarrow0}\, \Big(
T_+(w\!+\!i\e) \:\eta\!\left(T^{-1}_-(w\!-\!i\e)\right)\!\Big) 
\;,\quad \mbox{for}\quad w\in\R\;,
\end{equation}
peels off the compact part of the factorization \Ref{rc} and thus
coincides with the values of the original field on the symmetry axis  
$r\!=\!0$:  
\begin{equation}\la{Mg}
\cM(w) ~=~ 
\cV(r\!=\!0,t\!=\!-w)\:\,\eta\!\left(\cV^{-1}(r\!=\!0,t\!=\!-w)\right)
~=~g(r\!=\!0,t\!=\!-w) \;.
\end{equation}
In particular, it is real
\begin{equation}\la{real}
\cM(w)=\overline{\cM(w)}\;,
\end{equation}
and satisfies
\begin{equation}\la{sym}
\cM(w)=\eta\left(\cM(w)\right)^{-1}\;.
\end{equation}
Vice versa, \Ref{MTT} can be interpreted as the essentially unique
(Riemann-Hilbert) factorization of $\cM$ into a product of matrices
holomorphic in the upper and the lower half of the complex $w$-plane
respectively.\footnote{The matrix $\cM(w)$ in fact coincides with the
  so-called monodromy matrix of $\cVh_1$, originally introduced in
  \ci{BreMai87}. It is related to the transformation behavior of
  $\cVh_1$ between the two sheets of $\g$ and may in particular be
  extracted from $\cVh_1$ already at finite $r$.} 

Equ.\ \Ref{Mg} provides a physical interpretation for the new
constants of motion. Having been defined as spatially nonlocal charges
for fixed $t$, they gain a definite localization in the $2d$ spacetime
at fixed $r$. Moreover, they contain the entire information about the
solution: Together with the fact that $(\p_r\cV)(r\!=\!0)=0$ which
follows from the equations of motion \Ref{eqm}, the values on the
symmetry axis $r\!=\!0$ allow to recover the field $\cV$ everywhere.
In some sense the initial values on a spacelike surface have been
transformed into initial values along a timelike surface. Thus, the
$T_\pm(w)$ build a complete set of constants of motion for this
classical sector of solutions regular on the symmetry axis.

It remains to calculate the Poisson structure of the $T_\pm(w)$. The
result \Ref{PBT} in this case simplifies considerably. The first four
terms become
\ben
\left[\frac{\O_\gh}{v-w}\;,\; \stackrel1{T}\!(v)\stackrel2{T}\!(w)
\right]\;,
\een
for arbitrary indices $\pm$ at the $T$'s. The next two terms contain
$\g$ depending on the left boundaries $x, x'$ and tend to distinct
fixed values. The arising ambiguity at $x\!=\!x'$ however is precisely
cancelled by the additional contributions from the brackets
$\{T(x,y,v),\cV(x')\}$ and $\{\cV(x),T(x',y',w)\}$ in the limit $x,
x'\rightarrow0$. Only the last two terms are sensitive to the choice
of indices $\pm$. In the limit $y, y'\!\rightarrow\!\8$, the
corresponding combinations of the form \Ref{amb} tend to $+1$ if
$\g(v)$ and $\g(w)$ lie in the same of the two regions $D_+$ and
$D_-$, whereas they tend to $-1$ otherwise (cf.\ \Ref{limits}).

The final result is the following Poisson structure:
\ba
\left\{\stackrel1{T_\pm}\!(v)\;,\,\stackrel2{T_\pm}\!(w)\right\}
&=& \left[\frac{\O_\gg}{v-w}, 
\stackrel1{T_\pm}\!(v)\stackrel2{T_\pm}\!(w)
\right]\;,\la{T1}\\
\left\{\stackrel1{T_\pm}\!(v)\;,\,\stackrel2{T_\mp}\!(w)\right\}
&=& \frac{\O_\gg}{v-w} \stackrel1{T_\pm}\!(v)\stackrel2{T_\mp}\!(w)
-\stackrel1{T_\pm}\!(v)
\stackrel2{T_\mp}\!(w)\frac{\O_\gg^\eta}{v-w}\;,
\la{T2}\ea
with $\O_\gg^\eta\equiv \O_\gh-\O_\gk$ obtained from $\O_\gg$ by
applying $\eta$ in one of the two spaces.  Equs.\ \Ref{T1} build two
semi-classical copies of the Yangian algebra that is well known from
other $2d$ field theories \ci{Bern91,BerLec93,BouSch96}. By
semi-classical we conventionally mean that the Poisson brackets
\Ref{T1} coincide with the commutator of the $\hbar$-graded Yangian
algebra in first order $\hbar$. The mixed relations \Ref{T2} appear
``twisted'' by the involution $\eta$ with respect to those coming from
the normal Yangian double.  Note that whereas \Ref{T1} remains regular
at coinciding arguments, \Ref{T2} becomes obviously singular at
$v\!=\!w$. However, since $T_+$ and $T_-$ are defined in distinct
domains, this singularity appears only in the limit on the real line
and thus with a well-defined $\i\e$-prescription.

The matrices $\cM(w)$ form a closed Poisson algebra:
\ba\la{PBM}
\left\{\stackrel1{\cM}\!(v)\;,\,\stackrel2{\cM}\!(w)\right\}
&=&  \frac{\O_\gg}{v-w} \stackrel1{\cM}\!(v)\stackrel2{\cM}\!(w)
+  \stackrel1{\cM}\!(v)\stackrel2{\cM}\!(w) \frac{\O_\gg}{v-w} \\
&& {} -
\stackrel1{\cM}\!(v)\frac{\O_\gg^\eta}{v-w}\stackrel2{\cM}\!(w)
- \stackrel2{\cM}\!(w)\frac{\O_\gg^\eta}{v-w}\stackrel1{\cM}\!(v) 
\;.\nn
\ea
The singularity at $v\!=\!w$ is understood in the principal value
sense. For consistency, it may be checked that \Ref{PBM} is indeed
compatible with the symmetry \Ref{sym} of $\cM$.

Summarizing, the variables $\cM(w)$ provide a new formulation of the
model based on the closed Poisson algebra \Ref{PBM} and a physical
interpretation according to \Ref{Mg}. The further study and
quantization of this structure will start from the factorization
\Ref{MTT} with the corresponding Poisson algebra \Ref{T1}, \Ref{T2}.

Let us finally compute the Poisson brackets between the Hamiltonian of
\Ref{H} with the new variables $T_\pm$. Though they are integrals of
motion they do not commute with $H$, since their definition makes
explicit use of the time $t\!=\!\rt$.
The standard formula
\begin{equation}\la{TX}
\left\{\stackrel1{T}\!(x,y,v)\:,\stackrel2{X}\right\}
=\int_x^y dx'\stackrel1{T}\!(x,x',v)\left\{
\stackrel1{L_1}\!(x',\g(x',v))\:,\stackrel2{X}\right\}
\stackrel1{T}\!(x',y,v) \;,
\end{equation}
for arbitrary $X$, allows to explicitly calculate the Poisson bracket
\begin{equation}\la{PBH}
\{T_\pm(w),H\} = \p_w T_\pm(w)\;.
\end{equation}
This result also follows from a simple reasoning: due to the form
\Ref{g} of the spectral parameter $\g$ and the definition \Ref{TT}, the
explicit time dependence of the $T_\pm(w)$ (the one which is not
governed by $H$) equals their $w$-dependence. Since they are integrals
of motion, their bracket with $H$ then takes the form \Ref{PBH}.

\section{Timelike Dilaton}

In this chapter we study the case of a globally timelike vector field
$\p_\m\r$, which allows to identify $\r$ with the time $t$.
Accordingly, $\rt$ now builds the spatial coordinate $x$. The
distinguished location $\r\!=\!0$ which has played the role of the
symmetry axis $r\!=\!0$ in the previous chapter becomes now the origin
$t\!=\!0$. With periodic spatial topology, this is the setting of the
so-called cosmological Gowdy-models \ci{Gowd74}.\footnote{See
  \ci{Husa96,Mena97} for a recent treatment of the Gowdy model in
  Ashtekar variables.}  We will however just treat the asymptotic case
$x\in]-\!\8,\8[$. The fundamental structures of the preceding chapter
reappear in this context from a somewhat different side. We keep the
technical issues rather briefly here since they have been discussed in
detail above.

Again the transition matrices $T(x,y,t,w)$ provide a solution of the
linear system:
\begin{equation}\la{V2}
\cVh_{2}(x,t,\g(w))\equiv T(-\8,x,t,w)\;,
\end{equation}
for $\g(w)\in D_+\cup D_-$ inside of the unit circle. It gives rise to
the integrals of motion
\begin{equation}\la{Ttl}
T(w)\equiv T(-\8,\8,t,w)\;.
\end{equation}
With \Ref{limits} and \Ref{PBT} it follows that they satisfy the
Poisson algebra
\begin{equation}
\left\{\stackrel1{T}\!(v)\;,\,\stackrel2{T}\!(w)\right\}
= \left[\frac{\O_\gg}{v-w}, \stackrel1{T}\!(v)\stackrel2{T}\!(w)
\right]\;.
\end{equation}
In contrast to the previous chapter this function is continuous in the
$w$-plane (since the branch-cut does not blow up in the limit
$x\rightarrow\8$ but rather moves along the real line).  Moreover,
they do not contain the complete information about the model. This may
be seen most easily for solutions of \Ref{eqm} which are regular at
the origin $t\!=\!0$. For these solutions \Ref{Ttl} can be calculated
in the limit $t\rightarrow0$ where it becomes trivial. Since it is
$t$-independent we arrive at
\begin{equation}\la{triv}
T(w)=I\;.
\end{equation}
Thus, additional integrals of motion are required. 
Define 
\begin{equation}\la{Mfin}
\cM(w)=\lim_{\e\rightarrow0}\, \bigg(
\cVh_{2}(x,t,\g(w+\i\e))\:
\eta\Big(\cVh^{-1}_{2}(x,t,\g(w-\i\e))\Big)\!\bigg)\;,
\end{equation}
which is independent of $x$ and $t$. This is the proper analogue of
\Ref{MTT}. In fact within the setting of the previous chapter
definition \Ref{Mfin} with $\cVh_{2}$ replaced by $\cVh_{1}$ equals
\Ref{MTT}: the latter one is obtained from the former one in the limit
$r\!\rightarrow\!\8$ while $\cM$ is independent of $r$. In the case of
timelike $\r$ on the other hand, \Ref{Mfin} cannot be expressed in
terms of \Ref{Ttl} (as is obvious from \Ref{triv}). This is due to the
fact that the limits $\e\!\rightarrow\!0$ and $x\!\rightarrow\!\8$ do
not interchange in \Ref{Mfin}, cf. Appendix \ref{spectral}.

For the solutions regular at the origin $t\!=\!0$, $\cM(w)$ can again
be calculated more explicitly. Since $\cM(w)$ is independent of $x$
and $t$, we may evaluate it at the branchpoint $x\equiv t\!-\!w$ and
subsequently perform the limit $t\rightarrow0$. It is important to
keep $x= t\!-\!w$ during the limit process, since otherwise $\cM(w)$
does not behave smoothly. This yields: 
\ba
\cM(w)&=& \lim_{t\rightarrow0, x=w-t}\left(
\cP\exp\int_{-\8}^x\!\!dx L_1(x,\g)\;\;\;
\cP\exp\int_{-\8}^x\!\!dx \;\eta\Big(L_1(x,\g)\Big)\right)\non
&=& \cV(x\!=\!-w,t\!=\!0)\;\eta\Big(\cV(x\!=\!-w,t\!=\!0)\Big)
\;.\la{Mtim}
\ea
Thus, $\cM(w)$ again coincides with the values of the physical field
$g$ at $\r\!=\!0$.

The set of $\cM(w)$ together with their canonical Poisson structure
\Ref{PBM} provides a proper set of fundamental variables to
parametrize the phase space of solutions of \Ref{eqm} regular at the
origin $t\!=\!0$.  Note, that the canonical formulation obviously
fails to cope with describing this truncated phase space: At $t\!=\!0$
this framework breaks down with the vanishing Lagrangian \Ref{L},
whereas at finite $t$ the condition of regularity at $t\!=\!0$ poses
highly nontrivial implicit relations between the canonical coordinates
and the momenta.

After some calculation, the general form of the Poisson brackets
\Ref{PBT} further yields the same Poisson algebra \Ref{PBM} as in the
previous chapter for the matrices $\cM(w)$. Via the Riemann-Hilbert
decomposition of $\cM$ discussed above one may then further implicitly
obtain the matrices $T_\pm$ with Poisson-structure \Ref{T1}, \Ref{T2}.
Thus, together with \Ref{Mtim} for solutions regular at
$\r\!=\!0$ the final situation appears rather similar to the previous
chapter. This further stresses the fundamental meaning of the Poisson
structure \Ref{PBM}.

Let us finally look at the results of this chapter from another very
intriguing point of view. In the setting with a spacelike dilaton
addressed in the previous chapter we could have derived a Poisson
structure not with respect to the time $t$ but with respect to the
radius $r$.\footnote{In a covariant theory this is a rather natural
  idea which has been discussed in particular to describe static
  settings \ci{CaAlFi96}. For the Schwarzschild black hole e.g.\ one
  might throw doubt upon the distinct role of time in the canonical
  formalism since $r$ and $t$ change their character being space-
  resp. timelike inside of the horizon.} The calculations of this
chapter show that these two Poisson structures of the same model
coincide for the values of the original fields on the symmetry axis
$r\!=\!0$. Since these initial values provide a complete set of
observables the symplectic structures are essentially equivalent. It
is then tempting to speculate about further exploiting the
fundamental structure \Ref{PBM} even in the case of a timelike
dimensional reduction, i.e.\  the reduction to stationary axisymmetric
spacetimes, where the canonical time is no longer present.

\section{The Geroch Group}
With the integrals of motion $T_\pm(w)$ identified in the previous
chapters, one can now study the symmetries they generate via their
adjoint action in the canonical Poisson structure. This yields an
explicit realization \ci{KorSam97a} of the Geroch group \ci{Gero72}
with the underlying Yangian algebra \Ref{T1}, \Ref{T2}. The
transformations which close into an affine algebra (the loop algebra
$\hat{\mathfrak{g}}$) do not preserve the symplectic structure. This
is a particular example of the Lie-Poisson action of dressing groups
generated by the transition matrices of integrable models
\ci{Seme85,BabBer92,Lu93}. In contrast to the general case, here the
transition matrices $T_\pm(w)$ themselves are integrals of motion.

Let $\L(w)\in\gg$ be an algebra-valued function which is regular along
the real $w$-axis and vanishes at $w\mapsto\8$. Choose a
path $\ell=\ell_+\cup\ell_-$ encircling the real $w$-axis, such that
$\ell_\pm\in H_\pm$ and $\L(w)$ is holomorphic inside the enclosed
area. Define 
\begin{equation}\la{symmetry}
S[\L] ~\equiv~\tr \left(
\int_{\ell_+}  T^{-1}_+(w)\,\L(w)\; \ad_{T_+(w)} \,dw 
+ \int_{\ell_-} T^{-1}_-(w)\,\L(w)\; \ad_{T_-(w)} \,dw \right)\;,
\end{equation}
where ``$\ad$'' denotes the adjoint action via the canonical Poisson
structure. Since \Ref{T1}, \Ref{T2} yield
\ben
\left\{\stackrel1{T_\pm}\!(v)\;,\,\stackrel2{\cM}\!(w)\right\}
~=~ \left(\O_\gg\stackrel2{\cM}\!(w)
- \stackrel2{\cM}\!(w)\O_\gg^\eta\right)
\frac{\stackrel1{T_\pm}\!(v)}{v-w}\;,
\een
we obtain the symmetry action:
\begin{equation}\la{actionM}
S[\L] \cM(w) ~=~ \L(w)\cM(w) - \cM(w)\eta(\L(w))\;.
\end{equation}
This is the known infinitesimal action of the Geroch group on the
matrix $\cM$ \ci{BreMai87}. Even though the symmetry action
on the fields is highly nonlinear (cf.\ \Ref{actionV} and \Ref{actionP}
below), on the axis $\r\!=\!0$ it linearizes to \Ref{actionM} and allows
explicit ``exponentiation'' to finite transformations. This in
particular shows transitivity of the Geroch group in the sector of
solutions regular on the axis $\r\!=\!0$.

The transformations \Ref{actionM} form the loop algebra
$\hat{\gg}$, as follows also directly after some calculation from
\Ref{symmetry}, \Ref{T1} and \Ref{T2}:
\begin{equation}\la{affine}
\Big[S[\L_1],S[\L_2]\Big] ~=~ S\Big[[\L_1,\L_2]\Big]\;.
\end{equation}

Using \Ref{TX} allows to explicitly calculate the action of
$S[\L]$ on the physical fields $\cV$:
\ba
S[\L]\, \cV(x) &=& \int_\ell dw \left(
\frac{2\g}{\r(1-\g^2)}\,\cV(x)
\left[\cVh^{-1}(x,w)\L(w)\cVh(x,w)\right]_\gk\right) \la{actionV}\\
&=& - \int_{\g(\ell)} \g\,d\g\,\left( \cV(x)
\left[\cVh^{-1}(x,w)\L(w)\cVh(x,w)\right]_\gk\right) \;,\nn
\ea
with the algebra projection $[.]_\gk$ corresponding to the
decomposition \Ref{hk} and where $\cVh$ should be replaced by $\cVh_1$
or $\cVh_2$ respectively, corresponding to the choice of the dilaton
$\rho$. The corresponding transformation of the currents $P_\pm\equiv
\ft12(P_0\pm P_1)$ reads:
\begin{equation}\la{actionP}
S[\L]\, P_\pm\!(x) ~=~ \int_\ell dw \left(\;
\left[\frac{2\g}{\r(1\pm\g)^2}\,\left[\cVh^{-1}\L\cVh\right]_\gh
 , P_\pm\!(x)\right]
\;\mp\;\frac{4\g^2\p_\pm\r\left[\cVh^{-1}\L\cVh\right]_\gk}
{\r^2(1\pm\g)^2(1-\g^2)} 
 \;\right) \;.
\end{equation}
Equivalent forms of the infinitesimal symmetry transformations of the
Geroch group have been stated in
\ci{HauErn80,WuGe83,ChaGe88,Nico91}. E.g.\ it is easy to check, that
the single symmetry transformations as they are made explicit in
\ci{Nico91} may be obtained from the closed form \Ref{actionP} by
means of a Taylor expansion around $w\!=\!\8$.   

{}From \Ref{PBH} we further directly obtain the action of $S[\L]$ on
the Hamiltonian $H$:
\be\la{actionH}
S[\L]\, H ~=~ 
\tr \left(
\int_{\ell_+}  \L(w)\; \p_wT_+T^{-1}_+(w) \,dw 
+ \int_{\ell_-} \L(w)\; \p_wT_-T^{-1}_-(w) \,dw \right)\;,
\ee
in accordance with the formula derived in \ci{Nico91}.

Definition \Ref{symmetry} explicitly shows that this action of the
Geroch group is not symplectic but Lie-Poisson, i.e.\ it does not
preserve the Poisson structure on the phase space but on the direct
product of the phase space with the symmetry group. Its role in the
quantum theory remains to be elaborated, see \ci{Lu93} for a general
discussion. In our model we could alternatively consider the pure
symplectic action of the generators $T_\pm(w)$ via Poisson bracket
since they are integrals of motion themselves. However, though they
certainly act symplectic this action allows neither explicit
exponentiation nor a closed form of the commutator algebra in contrast
to \Ref{actionM} and \Ref{affine}.

\section{Quantization: A Twisted Yangian Double at the Critical Level} 

So far, we have achieved a complete reformulation of the classical
model \Ref{L} in terms of the transition matrices as new fundamental
variables providing a complete set of integrals of motion. This
formulation reveals integrability and the classical symmetries in a
beautiful way. We can now proceed with canonical quantization of the
Poisson algebra derived in the previous chapters.
For the sequel we specify the algebra $\gg$ to be $\mathfrak{sl}(N)$,
i.e.\ $\eta(X)=-X^t$ and $\gh=\mathfrak{so}(N)$. 
Let us recall the classical algebra of integrals of motion \Ref{T1},
\Ref{T2}. For $\gg=\mathfrak{sl}(N)$ it is
$\O_{\mathfrak{sl}(N)}=\Pi_N\!-\!\frac1{N}I$ with the $N^2\!\times\!N^2$
permutation operator $\Pi_N$:
\ben
\left(\Pi_N\right)^{ab,cd} = \d^{ad}\d^{bc} \;.
\een
Accordingly we define its twisted analogue $\Pi_N^\eta$ by
\ben
\left(\Pi_N^\eta\right)^{ab,cd} ~\equiv~ 
\left(-\Pi_N^{\tilde{t}}+\ft2{N}I\right)^{ab,cd} ~\equiv~  
-\d^{ac}\d^{bd} + \ft2{N}\d^{ab}\d^{cd} \;.\;
\footnotemark
\een
\footnotetext{The transposition $\Pi_N^{\tilde{t}}$ here amounts to
transposing just one of the two spaces in which $\Pi_N$ lives.}
The Poisson algebra \Ref{T1}, \Ref{T2} then takes the form:
\ba
\left\{\stackrel1{T_\pm}\!(v)\;,\,\stackrel2{T_\pm}\!(w)\right\}
&=& \left[\frac{\Pi_N}{v-w},
  \stackrel1{T_\pm}\!(v)\stackrel2{T_\pm}\!(w) \right]\;,\la{TT1}\\
\left\{\stackrel1{T_\pm}\!(v)\;,\,\stackrel2{T_\mp}\!(w)\right\}
&=& \frac{\Pi_N}{v-w} \stackrel1{T_\pm}\!(v)\stackrel2{T_\mp}\!(w)
- \stackrel1{T_\pm}\!(v)\stackrel2{T_\mp}\!(w)\frac{\Pi_N^\eta}{v-w}\;,
\la{TT2}
\ea
The $T_\pm(w)$ are related by complex conjugation \Ref{Tpm} and
further restricted by $T_\pm(w)\in\GG$: 
\be\la{det}
\det T_\pm(w)=1\;, 
\ee
and the additional property \Ref{sym}:
\be\la{sym1}
\cM(w)= T_+(w) T_-^t(w) = T_-(w) T_+^t(w) = \cM^t(w)\;.
\ee

Quantization of the model in terms of these variables now amounts to
replacing \Ref{TT1}, \Ref{TT2} by corresponding commutator relations
of an $\hbar$-graded algebra, such that these relations are compatible
with the quantum analogues of \Ref{det} and \Ref{sym1}. 
It admits the
following essentially unique solution:\footnote{For simplicity we use
  the same notation for the classical fields and the quantum
  operators. This should not lead to any confusion.}    

{\em The quantization of the presented model for
$\gg\!=\!\mathfrak{sl}(N)$ is given by the $*$-algebra generated by
the matrix entries of $N\times N$ matrices $T_\pm(w)$ subject to the
exchange relations
\ba
R(v\!-\!w)\stackrel1{T_\pm}\!(v)\stackrel2{T_\pm}\!(w) &=&
\stackrel2{T_\pm}\!(w)\stackrel1{T_\pm}\!(v) R(v\!-\!w) \;,\la{YN1}\\
R(v\!-\!w\!-\!\i\hbar)\stackrel1{T_-}\!(v)\stackrel2{T_+}\!(w) &=&
\stackrel2{T_+}\!(w)\stackrel1{T_-}\!(v) 
R^\eta(v\!-\!w\!+\!\ft2{N}\i\hbar)\,\chi(v\!-\!w)\la{YN2}\;,
\ea
with
\begin{equation}\la{RRchi}
R(v)\equiv vI-\i\hbar\Pi_N\;,\quad 
R^\eta(v)\equiv vI-\i\hbar\Pi_N^\eta\;,
\quad
\chi(v)\equiv
\frac{\G\Big(\ft{-\i\hbar-v}{N\i\hbar}\Big)
\G\Big(\ft{(N\!+\!2)\i\hbar-v}{N\i\hbar}\Big)}
{\G\Big(\ft{-v}{N\i\hbar}\Big)
\G\Big(\ft{(N\!+\!1)\i\hbar-v}{N\i\hbar}\Big)}\;.
\ee
The condition of unit determinant \Ref{det} is replaced by the quantum
determinant
\ba
{\rm qdet} T_\pm(w) &\equiv& 
\sum_{\s\in\mathfrak{S}_N} {\rm sgn}(\s) 
T_\pm^{1\s(1)}(w\!-\!(N\!\!-\!\!1)\i\hbar)
T_\pm^{2\s(2)}(w\!-\!(N\!\!-\!\!2)\i\hbar)\dots
T_\pm^{N\s(N)}(w) \non
&=& 1 \;,\la{qdetN}
\ea
and the quantum form of the symmetry \Ref{sym1} is given by
\begin{equation}\la{qsymN}
\cM(w)~\equiv~ T_+(w)T_-^t(w) ~=~ \left(T_+(w)T_-^t(w)\right)^t\;,
\end{equation}
where transposition here simply refers to the $N\times N$ matrix
entries. The $*$-operation is defined by
\begin{equation}\la{star}
T_+(w)^* \equiv T_-(\bar{w})\;,
\end{equation}
and builds a conjugate-linear anti-multiplicative automorphism of the
algebra \Ref{YN1}--\Ref{qsymN}. }
\bigskip

Denote by $Y_\pm$ the algebra generated by the $T_\pm(w)$ respectively
with exchange relations \Ref{YN1}. These are two copies of the
well-known Yangian algebra \cite{Drin85} which provide the unique
quantization of the Poisson algebras given by \Ref{TT1}. Compatibility
with associativity is equivalent to the Yang-Baxter equation for $R$:
\begin{equation}\la{YB}
R_{12}(u\!-\!v)R_{13}(u\!-\!w)R_{23}(v\!-\!w) =
R_{23}(v\!-\!w)R_{13}(u\!-\!w)R_{12}(u\!-\!v) 
\;, 
\end{equation}
The corresponding compatibility of the mixed relations \Ref{YN2} with
associativity is equivalent to a modified (twisted) Yang-Baxter
equation for $R^\eta$: 
\begin{equation}\la{TYB}
R^\eta_{12}(u\!-\!v)R^\eta_{13}(u\!-\!w)R_{23}(v\!-\!w) =
R_{23}(v\!-\!w)R^\eta_{13}(u\!-\!w)R^\eta_{12}(u\!-\!v) \;.
\end{equation}
Validity of  this equation follows from the relation 
\be\la{retrel}
R^\eta(v\!+\!\ft2{N}\i\hbar) = -\Big(R(-v)\Big)^{\tilde{t}}\;,
\ee
and \Ref{YB} by applying transposition in the first space. Moreover,
\Ref{TYB} remains valid under a shift of the argument in $R^\eta(v)$
as well as under rescaling it with a factor $\chi$. Hence, whereas
the exchange relations for $Y_\pm$ are uniquely given by \Ref{YN1}
\ci{Drin85}, the most general ansatz for the mixed exchange relations
\Ref{YN2} is 
\begin{equation}\la{Y2g}
R(v\!-\!w+c_1\i\hbar)\stackrel1{T_-}\!(v)\stackrel2{T_+}\!(w) ~=~
\stackrel2{T_+}\!(w)\stackrel1{T_-}\!(v)
R^\eta(v\!-\!w+c_2\i\hbar) \,\chi(v\!-\!w) \;.
\end{equation}
The resulting algebra must respect the symmetry \Ref{qsymN} of
$\cM(w)$. More precisely we demand that
\be\la{relid}
(Y_+\oplus Y_-)\,\cI = \cI\,(Y_+\oplus Y_-) \;,
\ee
where $\cI\subset\cU(Y_+\oplus Y_-)$ is spanned by the antisymmetric
matrix entries of $\cM$. Relation \Ref{relid} ensures that the
antisymmetry of $\cM$ may be consistently imposed without implying
further relations. Equ.\ \Ref{relid} is not influenced by the
choice of $\chi$ but uniquely determines the values of the central
extensions $c_j$ in \Ref{Y2g} to be
\ben
c_1=-1\;,\quad c_2=\ft2{N} \;.
\een
At these values the exchange relations between $T_\pm$ and $\cM$ may be
written in the closed form
\ba
R(v\!-\!w\!-\!\i\hbar)\stackrel1{T_-}\!(v)\stackrel2{\cM}\!(w)
&=& 
\stackrel2{\cM}\!(w)R^\eta(v\!-\!w\!+\!\ft2{N}\i\hbar)
\stackrel1{T_-}\!(v)\,\chi(v\!-\!w) \la{TMR}\\
R(v\!-\!w)\stackrel1{T_+}\!(v)\stackrel2{\cM}\!(w)
&=& 
\stackrel2{\cM}\!(w)R^\eta(v\!-\!w\!+\!(1\!+\!\ft2{N})\i\hbar)
\stackrel1{T_+}\!(v)\,\chi(v\!-\!w)\;,\nn
\ea
and indeed imply \Ref{relid}.

The factor $\chi(v)$ in \Ref{YN2} is finally fixed from the
requirement that the quantum determinants from \Ref{qdetN} commute
with everything such that the relations \Ref{qdetN} are consistent
with the algebra multiplication. It is known \ci{IzeKor81,KulSkl82}
that the $\mbox{qdet}T_\pm$ span the center of $Y_\pm$ respectively,
thus $\chi(v)$ must ensure that they also commute with $Y_\mp$. An
essential identity for this calculation is \ci{MoNaOl96}
\ba
\mbox{qdet}T_\pm(w) A_N &=& 
A_N \stackrel1{T_\pm}\!(w)\stackrel2{T_\pm}\!(w\!-\!\i\hbar)\dots
\stackrel{N}{T_\pm}\!(w\!-\!(N\!-\!1)\i\hbar)\non
&=&
\stackrel{N}{T_\pm}\!(w\!-\!(N\!-\!1)\i\hbar)\dots
\stackrel2{T_\pm}\!(w\!-\!\i\hbar)\stackrel1{T_\pm}\!(w) A_N \;,\nn
\ea
where $A_N$ denotes the antisymmetrizer in the $N$ auxiliary spaces.
Modifying the calculation for the Yangian \ci{MoNaOl96} by
using our mixed relations \Ref{YN2} finally leads to 
\ben
A_N R'_{01}\dots R'_{0N} A_N \,\stackrel0{T_-}\!(v) \,
\mbox{qdet}T_+(w) A_N =
\mbox{qdet}T_+(w) A_N \stackrel0{T_-}\!(v) 
A_N R''_{01}\dots R''_{0N} A_N\:,
\een
with
\ben
R'_{0j}\equiv R_{0j}(v\!-\!w\!+(j\!-\!2)\i\hbar)\;,\quad
R''_{0j}\equiv R^\eta_{0j}(v\!-\!w\!+(j\!+\!\ft2{N}\!-\!1)\i\hbar)\,
\chi(v\!-\!w\!+\!(j\!-\!1)\i\hbar)\;.
\een
Now
\ben
A_N R'_{01}\dots R'_{0N} A_N =
\frac{v\!-\!w\!-\!2\i\hbar}{v\!-\!w\!-\!\i\hbar}\,
A_N\;, \footnotemark 
\een
\footnotetext{We thank A.\ Molev for pointing this out.}
which together with \Ref{retrel} implies
\ben
A_N R''_{01}\dots R''_{0N} A_N =
\frac{w\!-\!v\!-\!N\i\hbar}{w\!-\!v\!-\!(N\!-\!1)\i\hbar}\;  
\prod_{j=1}^N \chi(v\!-\!w\!+\!(j\!-\!1)\i\hbar)\;A_N\;.
\een
Combining these equations yields the functional equation for $\chi(v)$
which is solved by \Ref{RRchi}. Uniqueness of this solution follows
from its normalization at $\ft{\i\hbar}{v}\rightarrow\!0_-$:
\ben
\chi(v)~\rightarrow~ 1-\ft{\i\hbar}{v}\left(1+\ft2{N}\right)\;,
\qquad\mbox{for}\quad \ft{\i\hbar}{v}\rightarrow0_-\;,
\een
which is required in order to obtain the correct classical limit
\Ref{TT2} from \Ref{YN2}.

It remains to check that the $*$-operation defined by \Ref{star} is
indeed a conjugate-linear anti-multiplicative automorphism of the
structure \Ref{YN1}--\Ref{qsymN}. Compatibility of \Ref{YN1} and
\Ref{YN2} with \Ref{star} obviously follows from
$R(\bar{u})=-\overline{R(-u)}$,
$R^\eta(\bar{u})=-\overline{R^\eta(-u)}$,
$\chi(\bar{u})=\overline{\chi(u)}$ and the fact that $R$ and $R^\eta$
are symmetric under permutation of the two spaces. Invariance of the
restriction of unit quantum determinant \Ref{qdetN} under $*$ follows
from 
\ba
\mbox{qdet}(T_\pm(w))^* &=&  
\sum_{\s\in\mathfrak{S}_N} {\rm sgn}(\s) 
T_\mp^{N\s(N)}(\bar{w})\dots 
T_\mp^{1\s(1)}(\bar{w}\!+\!(N\!\!-\!\!1)\i\hbar) \non
&=& \mbox{qdet}(T_\mp(\bar{w}\!+\!(N\!\!-\!\!1)\i\hbar))\;,\nn
\ea
where for the second identity we have employed one of the many
properties of the quantum determinant \ci{MoNaOl96}. Finally,
compatibility of the symmetry relation \Ref{qsymN} with $*$ follows
directly from invariance of this relation under $*$:
\ben
\left(T_+(w)T_-^t(w)\right)^* ~=~ 
\left(T_+(w)T_-^t(w)\right)^t ~=~ T_+(w)T_-^t(w) 
\qquad\mbox{for } w\in\R\;.
\een
\qed

The algebra \Ref{YN1}--\Ref{qsymN} is a new structure which in fact
brings together some concepts which arose within the last years in the
theory of quantum groups. The exchange relations \Ref{YN1} define the
$\mathfrak{gl}(N)$ Yangian \ci{FaSkTa79,Drin85}. The definition of the
quantum determinant \Ref{qdetN} is known from the $\mathfrak{sl}(N)$
Yangian \ci{IzeKor81,KulSkl82,MoNaOl96}.
A central extension of the type appearing in the mixed relations
\Ref{YN2} has been introduced for quantum affine algebras in
\ci{ResSem90} and explicitly for the Yangian double in
\ci{Khor96,IohKoh96}. Its value here is uniquely fixed from the
requirement of compatibility with \Ref{qsymN}. From the abstract point
of view the central extension takes the critical value at which
\Ref{relid} holds, i.e.\ any representation of the algebra \Ref{YN1},
\Ref{YN2} factorizes over $\cI$. The normal (untwisted) Yangian double
has a critical value of the central extension at which it possesses an
infinite dimensional center \ci{ResSem90}. As we shall discuss in the
next chapter, for $N\!=\!2$ the algebra \Ref{YN1}--\Ref{qdetN} is in
fact isomorphic to the normal centrally extended Yangian double at
critical level.

The essential new ingredient of \Ref{YN1}, \Ref{YN2} is the
twist\footnote{We are aware that the notation of ``twist'' has been
  introduced in several contexts for quantum groups in general and
  even for the Yangians in particular. In Ref.\ \ci{Olsh92} e.g.\ the
  ``twisted Yangian'' denotes the Yangian for the algebra
  $\mathfrak{so}(N)$. However, we hope that our notation here
  will not cause further confusion.} $\eta$ in the mixed relations which
already appeared in the classical Poisson algebra. It is basically
this peculiarity which requires a new representation theory to be
developed. 

\begin{Remark}\rm
The symmetry property \Ref{qsymN} together with the definition
of the $*$-map guarantees that the object $\cM(w)$ 
is symmetric and invariant under $*$. In a unitary representation it
will thus form a self-adjoint operator. As such it is the natural
quantum object that according to \Ref{Mg} underlies the original
classical field on the symmetry axis. It satisfies closed exchange
relations 
\begin{equation}\la{YM}
R(v\!-\!w)\!\stackrel1{\cM}\!\!(v)
R^\eta(w\!-\!v\!+\!(1\!+\!\ft2{N})\i\hbar)\!\stackrel2{\cM}\!\!(w)
\ee
\ben\hspace{12em}=~
\stackrel2{\cM}\!\!(w)
R^\eta(v\!-\!w\!+\!(1\!+\!\ft2{N})\i\hbar)\!
\stackrel1{\cM}\!\!(v)R(w\!-\!v)\frac{\chi(v\!-\!w)}{\chi(w\!-\!v)}\;,
\een
which may be viewed as the quantization of \Ref{PBM}.
\end{Remark}

\mathversion{bold}
\section{$\gg=\mathfrak{sl}(2)$}
\mathversion{normal}
To illustrate the formulas of the preceding chapter we will now
discuss the particular case $\gg=\mathfrak{sl}(2)$. This is the model
which describes the two Killing vector field reduction of pure $4d$
Einstein gravity and correspondingly already deserves strong interest
from the point of view of quantum gravity. The corresponding quantum
model has been introduced in \ci{KorSam97b}. Remarkably in this case
there is an algebra isomorphism between our twisted and the normal
Yangian double, however this is no $*$-algebra isomorphism.

The exchange relations here read
\ba
R(v\!-\!w)\stackrel1{T_\pm}\!(v)\stackrel2{T_\pm}\!(w) &=&
\stackrel2{T_\pm}\!(w)\stackrel1{T_\pm}\!(v) R(v\!-\!w) \;,\la{Y1}\\
R(v\!-\!w-\i\hbar)\stackrel1{T_-}\!(v)\stackrel2{T_+}\!(w) &=&
\stackrel2{T_+}\!(w)\stackrel1{T_-}\!(v) 
R^\eta(v\!-\!w+\i\hbar)\,\chi(v\!-\!w)\la{Y2}\;,
\ea
with $R$ and $R^\eta$ from \Ref{RRchi}, where the permutation operator
$\Pi$ and its twisted analogue $\Pi^\eta$ are given by
\ben
\Pi~\equiv~
\pmatrix{ 1& 0& 0& 0 \cr  0& 0& 1& 0 \cr  0& 1& 0& 0 \cr  0& 0& 0& 1
  \cr}\;,\qquad \Pi^\eta~\equiv~I-\Pi^{\tilde{t}}~\equiv~
\pmatrix{0&0&0&\!\!\!\!\!-1 \cr 0&1&0&0 \cr 0&0&1&0 \cr
  \!\!-1&\;0&\;0&\;0 
  \cr}\;.  
\een
Moreover, $\chi$ may be evaluated from \Ref{RRchi}
\ben
\chi(v)=\frac{v(v-2\i\hbar)}{(v-\i\hbar)(v+\i\hbar)} \;.
\een
The quantum determinant is given by
\begin{equation}\la{qdet}
{\rm qdet} T_\pm(w) ~\equiv~ 
T_\pm^{11}(w\!-\!\i\hbar)T_\pm^{22}(w)-
T_\pm^{12}(w\!-\!\i\hbar)T_\pm^{21}(w) ~=~1 \;;
\end{equation}
the matrix product 
\be\la{M2}
\cM(w) \equiv T_+(w)T_-^t(w) = \cM(w)^t \;,
\ee
is symmetric under transposition and satisfies \Ref{YM}.
 
The particular case $N\!=\!2$ is distinguished from the higher $N$
already on the classical level by the fact that the involution $\eta$
is an inner automorphism\footnote{In contrast, for $N\!>\!2$ the
  involution $\eta(X)=-X^t$ is the only outer automorphism of
  $\mathfrak{sl}(N)$, related to reflection of the Dynkin diagram.}
generated by conjugation with 
\ben
\s_2 = \pmatrix{\:0&\i\cr \!-\i&0} \;.
\een
This allows to ``retwist'' the mixed relations \Ref{Y2} by the
following transformation:
\be
\widetilde{T}_+(w)\equiv T_+(w)\s_2 \;,\qquad
\widetilde{T}_-(w)\equiv T_-(w)\;.
\ee
These retwisted generators satisfy the exchange relations of the
normal Yangian double:
\ba
R(v\!-\!w)\stackrel1{\widetilde{T}_\pm}\!(v)
\stackrel2{\widetilde{T}_\pm}\!(w) &=&
\stackrel2{\widetilde{T}_\pm}\!(w)
\stackrel1{\widetilde{T}_\pm}\!(v) R(v\!-\!w) \;,\la{Yd1}\\
R(v\!-\!w-\i\hbar)\stackrel1{\widetilde{T}_-}\!(v)
\stackrel2{\widetilde{T}_+}\!(w) &=&
\stackrel2{\widetilde{T}_+}\!(w)\stackrel1{\widetilde{T}_-}\!(v) 
R(v\!-\!w+\i\hbar)\:\chi(v\!-\!w)\la{Yd2}\;,
\ea
at the critical level $k\!=\!-2$. At this level the center of the
Yangian double becomes infinite-dimensional and is generated by the
trace of the quantum current \ci{ResSem90}
\be\la{qc}
L(w)\equiv \left[\widetilde{T}_+(w) 
\widetilde{T}^{-1}_-(w)\right] \;.
\ee
Evaluating this in terms of our matrix $\cM(w)$ from \Ref{M2} yields
\be\la{qch}
\tr\, L(w) = \cM^{12}(w) - \cM^{21}(w) \;.
\ee
The central extension of our structure was precisely determined by the
requirement \Ref{relid}. Since for $N\!=\!2$ the subspace $\cI$ is
one-dimensional, \Ref{relid} and an explicit calculation shows that it
even lies in the center of the algebra \Ref{Y1}--\Ref{Y2}. Here we see
the complete agreement with the normal Yangian double at critical
level. We have thus equivalence of the twisted structure
\Ref{Y1}--\Ref{Y2} with the untwisted \Ref{Yd1}--\Ref{Yd2}, however
supplied with a somewhat peculiar $*$-structure:
\ben
\widetilde{T}_+(w)^* = \widetilde{T}_-(\bar{w}) \s_2 \;.
\een
For higher $N$ this equivalence does not hold. Neither is there an
algebra isomorphism between \Ref{YN1}, \Ref{YN2} and the normal Yangian
double, nor does a center emerge at our critical level, rather
criticality is expressed by \Ref{relid}. 

\begin{Remark}\rm
Recall that classically the decomposition \Ref{MTT} provided the
Riemann-Hilbert decomposition of $\cM$ into a product of matrices
holomorphic in the upper resp.\ the lower half of the complex plane. It
is rather interesting to see how this classical holomorphy condition
changes in the quantum case.  Namely, assuming that the operators
$T_\pm(w)$ still act holomorphically in $H_\pm$ respectively, choose
some $w\!\in\!H_+$ near the real axis and integrate the mixed exchange
relation \Ref{Y2} in $v$ along some closed path in $H_-$ which
encircles $(w\!-\!\i\hbar)$. Due to the nonvanishing r.h.s.\ this leads
to a contradiction caused by the quantum effect of shifting the
singularity in \Ref{Y2} by an amount of $\i\hbar$. Thus the action of
$T_-(v)$ can not be holomorphic in the entire $H_-$; this argument
shows that in fact it can not be represented holomorphically in any
sub-domain of the strip $-\hbar <\Im v < 0$. In the same way it
follows that the action of $T_+(v)$ can be holomorphic only in the
domain $\Im v > \hbar$.

We come to an interesting observation: consistent quantization of the
classical Riemann-Hilbert problem leads to a kind of ``quantum
Riemann-Hilbert problem'', where the contour itself (i.e. the real
line) spreads out; the domains of analyticity of $T_\pm$ in the
quantum model are $H_\pm^\hbar$, i.e. $H_\pm$ without a small strip
along the real line. This does not imply that $T_\pm$ can not be
defined inside of the strip $|\Im v|<\hbar$; it just means that they
cannot act holomorphically inside. Therefore, it still makes sense to
define the matrix $\cM(u)$, however, as a quantum operator it looses
its classical meaning as a patching matrix of the Riemann-Hilbert
problem.
\end{Remark}

\begin{Remark}\rm
For explicit calculations it is sometimes useful to express the
exchange relations \Ref{Y1}, \Ref{Y2} in matrix components
$T_\pm^{ab}(w)$. The mixed relations \Ref{Y2} e.g.\ may equivalently be
written as
\ba
\left(1-\frac{(\i\hbar)^2}{(v\!-\!w)^2}\right)
T_-^{ab}(v) T_+^{cd}(w) &=& \left(1-\frac{\i\hbar}{(v\!-\!w)}\right)
T_+^{cd}(w)T_-^{ab}(v) 
\la{compo}\\
&&\hspace{-3em}{}+\frac{\i\hbar}{(v\!-\!w)}
\Big(T_+^{ad}(w)T_-^{cb}(v)+\d^{bd}\,T_+^{cm}(w)T_-^{am}(v)\Big)\non 
&&\hspace{-3em}{} +\frac{(\i\hbar)^2}{(v\!-\!w)^2}\:\d^{bd}
\Big(T_+^{am}(w)T_-^{cm}(v)-T_+^{cm}(w)T_-^{am}(v)\Big)\;.\nn
\ea
Interpreting the matrix entries of the $T_\pm$ as creation and
annihilation operators respectively, the r.h.s.\ of \Ref{compo} can be
viewed as sort of normal ordering.
\end{Remark}

\section{Outlook}

We have given a complete reformulation of the classical models in
terms of the transition matrices of the associated linear system. In
contrast to the situation in general integrable models, here the
transition matrices themselves are integrals of motion. Moreover they
contain a complete set of conserved charges related to the values of
the physical fields on the axis $\r\!=\!0$.  The Poisson algebra of
these matrices has been shown to form a semi-classical version of the
Yangian double modified by appearance of a twist by the coset
involution $\eta$. The transitive action of the Geroch group becomes
manifest and rather transparent in this picture as the Lie-Poisson
action generated by the transition matrices. This classical picture
has been established for an arbitrary semisimple Lie algebra $\gg$
underlying the $\Coset$ coset $\s$-model.

Quantization for $\gg\!=\!\mathfrak{sl}(N)$ led to a twisted
Yangian double with central extension where the exact value of the
central extension is uniquely determined from consistency, more
precisely from compatibility of the structure with the symmetry of the
matrix $\cM$.  For $\gg\!=\!\mathfrak{sl}(2)$ the structure is in fact
isomorphic (but not $*$-isomorphic) to the centrally extended Yangian
double at the critical level with infinite-dimensional center.

Continuation of the programme is straightforward to outline. The
representation theory of the algebra \Ref{YN1}--\Ref{qsymN} must be
studied. So far only the finite-dimensional representations of the
normal Yangian are completely understood and classified \ci{ChaPre94}.
These results might serve as basic tools to support the first steps in
exploring the relevant infinite-dimensional representations of our
algebra. The hope is certainly that the requirement of unitarity with
respect to the $*$-structure \Ref{star} will strongly restrict the
choice of representations. In \ci{KorSam97b} we have suggested a
particular Fock space type representation where inspired by the linear
truncation of the model the two Yangian halves of \Ref{YN1} act as
creation and annihilation operators respectively. However, unitarity
of this representation is not obvious, so eventually one might have to
face states of negative norm; this remains to be investigated.

Once the set of possible representations has been identified and
hopefully been brought to a minimum the next goal is the construction
of coherent states in this framework. These states should exhibit
minimal quantum fluctuations around given classical solutions.  With
them at hand one would finally be in position to study in detail how
quantization affects the known classical solutions of gravity. In
particular, this might shed some light onto the discussion about
existence and properties of suited coherent states in the truncation
of the model to collinearly polarized gravitational waves
\ci{Asht96,GamPul97}.  The quantum analogue of the Geroch group is
supposed to play the key role of a spectrum generating group,
i.e.\ in accordance with the classical picture it should act
transitively among the coherent states. It may be possible to
shortcut the explicit construction of representation and coherent
states by properly understanding the quantization \ci{Lu93} of this
Lie-Poisson action that we have described in the classical picture.

Further open problems remain. At this stage we do not know the
explicit link of the canonical approach adopted here to the
isomonodromic quantization proposed in \ci{KorNic96,KorSam96} for the
same model. Although related quantum group structures appear, the
isomonodromic framework is formulated in terms of different
observables, which makes the comparison even on the classical
level rather nontrivial.    

Since the classical picture is already formulated for an arbitrary
semisimple Lie-algebra $\gg$ there remains the obvious task to
elaborate the quantization for higher-dimensional coset spaces. As
mentioned several times by now this corresponds to the models which
descend from dimensional reduction of matter-coupled gravities and
supergravities.  The study of their quantization so far suffers
simply from the fact that the theory of Yangians associated to the
exceptional groups e.g.\ is still strongly underdeveloped --- not to
mention their representation theory. To describe the corresponding
reduction of maximally extended $N\!=\!8$ supergravity \ci{Juli83} one
would have to construct the related possibly centrally extended
Yangian double of $E_{8(+8)}$ with a twist characterizing the maximal
compact subgroup $SO(16)$.

Another highly interesting generalization would be the extension of
this framework to a dimensional reduction which includes a timelike
Killing vector field. At present it seems rather subtle to rigorously
establish a canonical framework in a sector of stationary solutions
where the canonical time-dependence has been dropped by hand. On the
other hand it is certainly this sector which contains the most
interesting physical solutions, in particular the black holes.
Justifying relevance of the fundamental structures obtained in this
paper within that context would open the doors to a profound
understanding of quantum black holes.

\section*{Acknowledgements}
We would like to thank A.\ Kitaev, H.\ Nicolai, V.\ Schomerus and M.\ 
Semenov-Tian-Shansky for helpful discussions at different stages of
this work. The work of D.\ K. was partially supported by DFG Contract
Ni 290/5-1. H.\ S. thanks Studienstiftung des deutschen Volkes for
support.

\begin{appendix}
\section{The Spectral Parameters}\la{spectral}
The variable spectral parameter $\g$ is a function of the constant
spectral parameter $w$ according to 
\begin{equation}\la{g}
\g(x,t,w) ~=~ \g\left(\frac{w+\rt}{\r}\right) ~=~ 
\frac1{\r}\left(w+\rt-\sqrt{(w+\rt)^2-\r^2}\;\right)\;.
\end{equation}
It satisfies the differential equations
\begin{equation}
\g^{-1}\p_\m\g  ~=~ \frac{1+\g^2}{1-\g^2}\,\r^{-1}\p_\m\r 
+\frac{2\g}{1-\g^2}\,\e_{\m\nu}\r^{-1}\p^\nu\r\;,
\end{equation}
as well as 
\begin{equation}
 \g^{-1}\p_w \g ~= -\frac{2\g}{\r(1-\g^2)}\;.
\end{equation}
The inverse formula reads
\begin{equation}
w=\ft12\r(\g+\ft1{\g})-\rt\;.
\end{equation}

The parameter $\g$ lives on the Riemann surface defined by
$\sqrt{(w+\rt+\r)(w+\rt-\r)}$, which is a twofold covering of the
complex $w$-plane with $x^\m$-dependent branch-cut. Transition between
the two sheets corresponds to $\g\rightarrow\frac1{\g}$.  The
branch-cut connects the points $w\!=\!-\rt\pm\r$ on the real $w$-axis,
which correspond to $\g(w\!=\!-\rt\!\pm\!\r)=\pm1$. The real
$w$ with $|w+\rt|<|\r|$ are mapped onto the unit circle $|\g|=1$. Real
$w$ with $|w+\rt|>|\r|$ are mapped onto the real $\g$-axis. Dividing
the $w$-plane into two regions $H_\pm$ and the $\g$-plane into four
regions $D_\pm, \Dt_\pm$ according to Fig. \ref{planes}, $D_\pm$ and
$\Dt_\pm$ lie over $H_\pm$ respectively. 
\begin{figure}[htbp]
  \begin{center}
    \leavevmode \input{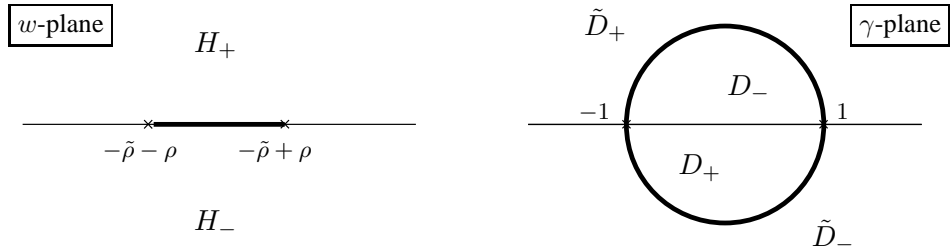}
  \end{center}
  \caption{The spectral parameter planes}
  \label{planes}
\end{figure}

In particular, it is important that  for fixed $w\notin\R$ and
continously varying $\r$ and $\rt$ the parameter $\g$ moves within a
fixed of these regions. The limits of these trajectories are given by
\begin{equation}\la{limits} \g(\r\!\rightarrow\!0) \rightarrow
\left\{\begin{array}{l}
    \!\!0\\
    \!\!\8
                  \end{array}\right. \;,\quad
\g(\r\!\rightarrow\!\8) \rightarrow \left\{\begin{array}{r} 
                               \i\\
                               -\i
                  \end{array}\right. \;,\quad
\g(\rt\!\rightarrow\!\pm\8) \rightarrow \left\{\begin{array}{l} 
                               \!\!0\\
                               \!\!\8 
                  \end{array}\right.\;. 
\end{equation}

Another useful formula for two spectral parameters $\g(x,t,v)$ and
$\g(x,t,w)$ at coinciding coordinates $x, t$ is:
\begin{equation}
v-w ~=~ \frac{\r}{2}\:\frac{(\g(v)\!-\!\g(w))\,(\g(v)\g(w)\!-\!1)}
{\g(v)\g(w)}\;,
\end{equation} 
which has e.g. been employed in the calculation of \Ref{PBT}.

\end{appendix}

\end{document}